\documentclass[aps,prb,notitlepage,superscriptaddress,%
longbibliography,twocolumn]{revtex4-2}

\usepackage{amsmath}
\usepackage{amssymb}
\usepackage{maybemath}
\usepackage[x11names]{xcolor}
\usepackage{graphicx}
\usepackage[caption=false]{subfig}
\usepackage{bm}
\usepackage{bbm}
\usepackage{soul}

\usepackage{siunitx}

\usepackage{dcolumn}
\newcolumntype{d}[1]{D{.}{.}{#1}}
\newcolumntype{.}{D{x}{}{9}}
\newcolumntype{,}{D{x}{}{5}}
\newcolumntype{;}{D{x}{}{19}}

\usepackage{booktabs}   
\usepackage{natbib}     
\newcommand{\ii}{\mathrm{i}}
\renewcommand{\Re}{{\mathrm{Re}~}}
\renewcommand{\Im}{{\mathrm{Im}~}}

\def\calD{{\mathcal D}}
\def\calE{{\mathcal E}}
\def\calF{{\mathcal F}}

\def\calN{{\mathcal N}}

\def\bbmA{{\mathbbm A}}
\def\bbmM{{\mathbbm M}}
\def\bbmR{{\mathbbm R}}
\def\bbmX{{\mathbbm X}}

\def\dd{{\mathrm d}}
\def\ii{{\mathrm i}}

\definecolor{garrosgreen}{rgb}{0.1, 0.4, 0.1}
\definecolor{dartmouthgreen}{rgb}{0.05, 0.5, 0.06}
\definecolor{duelferred}{rgb}{0.7, 0.2, 0.1}
\definecolor{cambridgeblue}{rgb}{0.1, 0.3, 1.0}
\definecolor{oxfordblue}{rgb}{0.05, 0.2, 0.7}

\begin{document}

\title{Coupled Oscillators and Dielectric Function}

\author{T. Das}
\affiliation{Department of Physics and LAMOR, Missouri University of Science and
Technology, Rolla, Missouri 65409, USA}

\author{C. A. Ullrich}
\affiliation{Department of Physics and Astronomy,
University of Missouri, Columbia, Missouri 65211, USA}

\author{U. D. Jentschura}
\affiliation{Department of Physics and LAMOR, Missouri University of Science and
Technology, Rolla, Missouri 65409, USA}
 
\begin{abstract}
A generalized Sellmeier model,
also referred to as the Lorentz--Dirac model,
has been used for the description
of the dielectric function of a
number of technologically important materials
in the literature.
This model represents the frequency-dependent
dielectric function as a sum over Green
functions of classical damped harmonic oscillators,
much in analogy with the functional form
used for the dynamic polarizability of an atom,
but with one important addition,
namely, a complex-valued oscillator
strength in the numerator.
Here, we show that this generalized
functional form can be justified based on the
response function of {\em coupled} damped oscillators.
The encountered analogies suggest
an explanation for the generally observed
success of the Lorentz--Dirac model in describing
the dielectric function of crystals of
consummate technological significance.
\end{abstract}

\maketitle

%
%
\section{Introduction}
\label{sec1}

There is no generally accepted universal
functional form for the frequency-dependent
dielectric function 
$\epsilon(\omega)$ of a solid. In order to illustrate
this statement, we observe that, for the various materials
considered in the seminal  Ref.~\cite{Pa1985},
vastly different functional forms have been
employed in the opening paragraphs
which precede the data listings contained
in Ref.~\cite{Pa1985}. Recently, very complicated
and involved functional forms have been explored
in Ref.~\cite{PSEtAl2009}, for calcium fluoride.

In order to obtain a more consistent picture,
it is useful to observe that the
dielectric function is (by definition)
proportional to the dielectric displacement inside the
material. Hence, the quantity
$[\epsilon(\omega) - 1]$ must be proportional
to the induced polarization $P$ 
(which equals the volume density of induced dipole moments, 
see Ref.~\cite{HaKo2009}). In some cases (not solids),
the functional form of the dielectric
function is known much better.
One example is a dilute gas,
where [see Eq.~(6.132) of Ref.~\cite{Je2017book}]
\begin{align}
\label{dieleceps}
\epsilon(\omega) =& \; 1 + \frac{N_V}{\epsilon_0} \; \alpha(\omega)
\nonumber\\[0.1133ex]
=& \; 1 + \frac{N_V}{\epsilon_0} \; \sum_n \frac{f_{n0}}{\hbar^2} \,
\frac{1}{\omega_n^2 - \omega^2 - \ii \, \gamma_n \, \omega} \,.
\end{align}
Here, $\alpha(\omega)$ is the dipole polarizability
of the gas atoms, $N_V$ is their number density, 
and $\epsilon_0$ is the vacuum permittivity ($\hbar$ is Planck's
unit of action).  The resonance frequencies
of the atomic transitions are denoted as $\omega_n$,
and their respective widths are $\gamma_n$
(for absolute clarity, we should note
that we shall use the term frequency as a synonym for the 
angular frequency throughout the article,
for brevity of notation).
The oscillator strengths are denoted as $f_{n0}$.
One notes that the sum over $n$, a priori,
includes all dipole-allowed transitions of the atom.

We have recently used a form analogous
to Eq.~\eqref{dieleceps} for the description
of the temperature-dependent
dielectric function of intrinsic silicon~\cite{MoEtAl2022},
\begin{equation}
\label{master}
\epsilon(T_\Delta, \omega) = 1 + \sum_{k=1}^{k_{\rm max}}
\frac{a_k(\omega_k^2 - \ii \gamma_k'\omega)}%
{\omega_k^2 - \omega^2 - \ii \omega \gamma_k} \,,
\end{equation}
where the oscillator strengths $f_{n0}$
of Eq.~\eqref{dieleceps}
are generalized to complex quantities.
For intrinsic silicon~\cite{MoEtAl2022}, we were able to achieve a
description of the available data
with two resonances ($k_{\rm max} = 2$).
In Appendix A.2 of Ref.~\cite{MoEtAl2022},
we argued that the presence of the
parameter $\gamma'_k$
in the numerator of Eq.~\eqref{master}
can be explained on the basis of radiative
reaction (Lorentz--Dirac equation).
One finds~\cite{MoEtAl2022,MoEtAl2024erratum}
that the temperature-dependence of the
parameters $a_k$ (amplitude), $\omega_k$
(resonance frequency), $\gamma_k$ (width)
and $\gamma'_k$ (radiative-reaction width)
is smooth and can be described by quadratic
polynomials in the variable $T_\Delta = (T-T_0)/T_0$,
where $T_0$ is room temperature (293 K).
This functional form is much simpler than,
{\em e.g.}, the ones employed in Ref.~\cite{PSEtAl2009}.

The model given in Eq.~\eqref{master}
was used in Eq.~(1) of Ref.~\cite{Ri1985} for
a description of the dielectric
function of rutile, in
Eq.~(4) of Ref.~\cite{Tr1985} for cubic thallium, and in
Eq.~(1) of Ref.~\cite{PaKh1985} for sodium nitrate.
Furthermore, it was used in
Ref.~\cite{Je2024multipole} for $\alpha$-quartz.
On the basis of the Lorentz--Dirac equation,
the parameter $\gamma'_k$ should be a positive quantity,
as explained in Appendix A.2 of Ref.~\cite{MoEtAl2022}.
This observation raises the question
why, for $\alpha$-quartz~\cite{Je2024multipole},
one finds negative $\gamma'_k$ parameters from
a fit of the dielectric function.
Here, we shall go a different route and
explore if one can find an analogy for the functional
form~\eqref{master} considering
coupled damped harmonic oscillators.

The motivation for this endeavor stems from the fact that the
denominator in Eq.~\eqref{master},
$\omega_k^2 - \omega^2 - \ii \omega \gamma_k$,
is characteristic of a damped harmonic
oscillator with resonance frequency $\omega_k$
and damping constant $\gamma_k$
(see Sec.~2.2 of Ref.~\cite{Je2017book}).
One might thus ask to which extent the
coupling of oscillators can be related
to the complex oscillator strength
[manifestly complex numerator
$a_k(\omega_k^2 - \ii \gamma_k'\omega)$]
in Eq.~\eqref{master}.
Our study thus goes beyond the textbook derivation
of the optical susceptibility based on
uncoupled classical oscillators \cite{HaKo2009}.

The paper is organized as follows.
In Sec.~\ref{sec2}, we consider
two coupled damped oscillators,
and we generalize the considerations to
three coupled oscillators in Sec.~\ref{sec3}.
The additional role of radiative reaction
is discussed in Sec.~\ref{sec4}.
In Sec.~\ref{sec5}, we discuss
the back-reaction of the dielectric response 
mediated by photon and phonon fields
onto the polarization, 
and analogies to coupled-oscillator models.
Conclusions are reserved for Sec.~\ref{sec6}.
Atomic units with $\epsilon = 1/(4 \pi)$,
$\hbar = |e| = 1$, and $c = 1/\alpha$
are employed (see Chap.~2 of Ref.~\cite{JeAd2022book}).
Here, $e$ is the electron charge and $|e|$ is its
modulus, while $\alpha \approx 1/137.036$ is the
fine-structure constant.
Some technical details are discussed in the Appendix.

%
%
\section{Coupled Damped Oscillators}
\label{sec2}

%
%
\subsection{Derivation of the Signal}

\begin{figure}
\includegraphics[width=\linewidth]{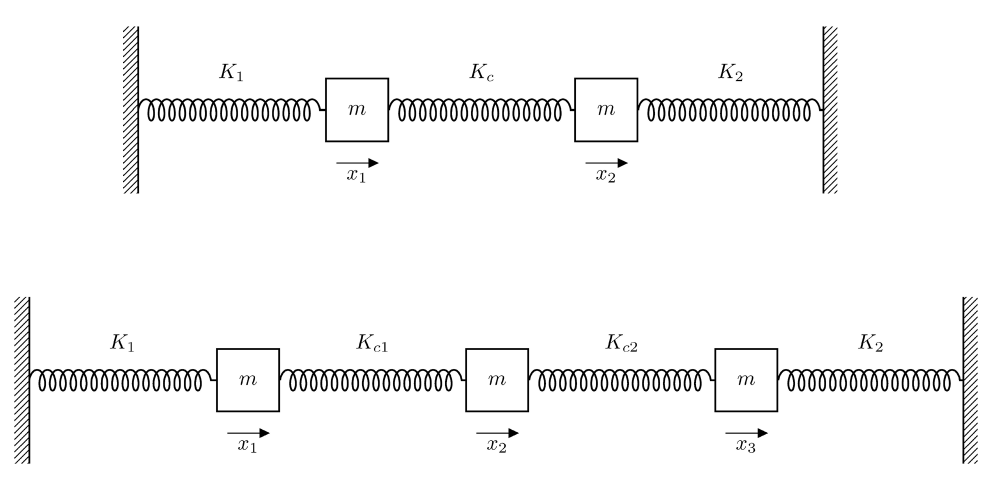}
\caption{\label{fig1} Schematic view of two
and three coupled oscillators (upper and lower panel, respectively).}
\end{figure}

In view of the mathematical analogies
of damped harmonic oscillators (and their Green functions),
analytic expressions for the polarizability
of atoms, and the dielectric function
of materials (see, {\em e.g.},
Sec.~6.4.3 of Ref.~\cite{Je2017book}),
it is indicated to consider the signal obtained
by driving coupled classical oscillators,
and its relation to  the mathematical
form of the Lorentz--Dirac model
given in Eq.~\eqref{master}.
We thus consider a system of two coupled
oscillators, with coordinates $x_1$ and $x_2$
and spring constants $K_1$ and $K_2$ for
the uncoupled oscillators, and $K_c$ for the
coupling (see Fig.~\ref{fig1}).

The following two questions will be investigated:
{\em (i)} Can the response of
coupled damped oscillators be
described to good numerical accuracy
by a functional form of the
Lorentz--Dirac model~\eqref{master},
that is, can the signal of
coupled damped oscillators alternatively
be described by uncoupled oscillators
with a complex-valued oscillator strength?
{\em (ii)} If the answer to {\em (i)} is affirmative,
are there cases where the radiative-reaction
parameter $\gamma'$ becomes negative
upon fitting the signal from the coupled resonances?

Based on Fig.~\ref{fig1}
one easily obtains the following dynamic equations:
\begin{align}
m \frac{\dd^2 x_1}{\dd t^2} =& \;
 - \Gamma_1 \frac{\dd x_1}{\dd t} - (K_1 + K_c) x_1 + K_c x_2 + F_1(t)  \,,
\\
m \frac{\dd^2 x_2}{\dd t^2} =& \;
 - \Gamma_2 \frac{\dd x_2}{\dd t} - (K_2 + K_c) x_2 + K_c x_1 + F_2(t)  \,.
\end{align}
Here, $m$ denotes the masses of the driven oscillators,
which we assume to be equal \cite{massfootnote}, and the $F_i$ with
$i=1,2$ are the force terms.
The relation of the spring constants to the (unperturbed)
resonance frequencies $\omega_{10}$ and $\omega_{20}$
is $K_1 = m (\omega_{10})^2 = m \, k_1$,
$K_2 = m (\omega_{20})^2 = m \, k_2$, and we also define
$K_c = m \omega_c^2 = m \, k_c$,
$\Gamma_1 = m \gamma_1$,
$\Gamma_2 = m \gamma_2$,
$F_1 = m \, f_1$, and $F_2 = m \, f_2$.
In the scaled variables, one obtains
\begin{subequations}
\label{eqsys}
\begin{align}
\frac{\dd^2 x_1}{\dd t^2} =& \;
 - \gamma_1 \frac{\dd x_1}{\dd t} - (k_1 + k_c) x_1 + k_c x_2 + f_1(t)  \,,
\\
\frac{\dd^2 x_2}{\dd t^2} =& \;
 - \gamma_2 \frac{\dd x_2}{\dd t} - (k_2 + k_c) x_2 + k_c x_1 + f_2(t)  \,.
\end{align}
\end{subequations}
Furthermore, we set $f_1 = a_1 \, E_1$
and $f_2 = a_2 \, E_2$, where the
$E_i$ describe the electric fields,
and the $a_i$ are proportional
to the number densities of the oscillators.
This amounts to a model of coupled oscillators
where the spring constants $k_1$ and $k_2$
are equal to the
squares of the unperturbed resonance
frequencies, $k_1 = (\omega_{10})^2$ and $k_2 = (\omega_{20})^2$.
The damping constants are $\gamma_1$ and $\gamma_2$,
and we assume a spring constant $k_c$ for the
coupling between the oscillators.

Various physical parameters such as the
charge of the driven system are set equal to unity
in the above system of equations, and we emphasize
that the coupled classical oscillators
can at best only be a {\em model problem} for the full
quantum-mechanical system under consideration.
We will discuss this point later in Sec.~\ref{sec5}.

In frequency space, after Fourier transformation,
one has
$\bbmM(\omega) \cdot \vec x(\omega) = \vec E(\omega)$,
\begin{multline}
\bbmM \cdot
\left( \begin{array}{cc}
x_1(\omega) \\ x_2(\omega) \end{array} \right)
= \bbmA \cdot
\left( \begin{array}{cc}
E_1(\omega) \\ E_2(\omega) \end{array} \right) \,,
\\[0.1133ex]
\bbmM = \left( \begin{array}{cc}
\bbmM_{11} & \bbmM_{12} \\ \bbmM_{21} & \bbmM_{22} \\
\end{array} \right) \,,
\qquad
\bbmA = \left( \begin{array}{cc}
\bbmA_{11} & \bbmA_{12} \\ \bbmA_{21} & \bbmA_{22} \\
\end{array} \right) \,.
\end{multline}
The elements of the matrices are easily obtained,
\begin{subequations}
\label{M}
\begin{align}
\bbmM_{11} =& \; k_1 + k_c - \ii \, \gamma_1 \, \omega - \omega^2 \,,
\\[0.1133ex]
\bbmM_{12} =& \; - k_c = \bbmM_{21} \,,
\\[0.1133ex]
\bbmM_{22} =& \; k_2 + k_c - \ii \gamma_2 \omega - \omega^2 \,,
\\[0.1133ex]
\bbmA_{11} =& \; a_1 \,, \qquad
\bbmA_{22} = a_2 \,, \qquad
\bbmA_{12} = \bbmA_{21} = 0 \,.
\end{align}
\end{subequations}
One then finds the inverse relation
\begin{equation}
\left( \begin{array}{cc}
x_1(\omega) \\ x_2(\omega) \end{array} \right) =
\bbmX \cdot
\left( \begin{array}{cc}
E_1(\omega) \\ E_2(\omega) \end{array} \right),
\end{equation}
where the elements of $\bbmX = \bbmM^{-1} \cdot \bbmA$ are
as follows,
\begin{subequations}
\begin{align}
\bbmX_{11} =& \frac{ N_1 }{ \calD }  \,, \qquad
\bbmX_{12} = \frac{a_2 k_c }{\calD } \,,
\\[0.1133ex]
\bbmX_{21} =& \; \frac{a_1 k_c }{\calD } \,, \qquad
\bbmX_{22} = \frac{ N_2 }{ \calD } \,,
\\[0.1133ex]
N_1 =& \; a_1 \,
(k_2 + k_c - \omega^2 - \ii \gamma_2 \omega ) \,,
\\[0.1133ex]
N_2 =& \; a_2 \,
(k_1 + k_c - \omega^2 - \ii \gamma_1 \omega ) \,.
\end{align}
\end{subequations}
The denominator $\calD = \det(\bbmM)$ is given by
\begin{equation}
\label{calD}
\calD =
(k_1 + k_c - \omega^2 - \ii \gamma_1 \omega)
(k_2 + k_c - \omega^2 - \ii \gamma_2 \omega)
- k_c^2 \,.
\end{equation}
For uncoupled oscillators ($k_c = 0$), one recovers the
structure of the Sellmeier model (with damping),
\begin{equation}
\left. \bbmX \right|_{k_c = 0} =
\left( \begin{array}{cc}
\dfrac{ a_1 }%
{ k_1 - \omega^2 - \ii \, \gamma_1 \omega} & 0 \\[2ex]
0 & \dfrac{ a_2 }%
{ k_2 - \omega^2 - \ii \, \gamma_2 \omega}
\end{array} \right) \,.
\end{equation}
For the response of the entire system,
one sums over the polarization densities
corresponding to the two oscillators.
Provided we assume both oscillators 
to be driven by the same electric field 
$E_1(\omega) = E_2(\omega) = E(\omega)$,
the electric susceptibility $\chi = 4 \pi P/E$ 
is obtained as a quantity proportional to 
\begin{align}
\label{P2}
\chi(\omega) \equiv & \;
{\rm Tr} \, \left[ \bbmX \cdot
\left( \begin{array}{c}
1 \\ 1
\end{array} \right) \right] =
\bbmX_{11} + \bbmX_{12} + \bbmX_{21} + \bbmX_{22}
\nonumber\\[0.1133ex]
=& \;
\frac{ N_1 + (a_1 + a_2) k_c + N_2 }{ \calD } \,.
\end{align}
Here, we have denoted the trace of a
two-component vector $\vec v$ (not a $2 \times 2$ matrix)
as the sum of its elements $v_1 + v_2$.
The deviation of the
relative dielectric function $\epsilon(\omega)$
from its vacuum value (unity)
is $\epsilon(\omega) - 1 = \chi(\omega)$.
In order to analyze the resonance
structure of the coupled system, it is
useful to observe that
\begin{equation}
\calD =
(\omega_{c1}^2 - \omega^2)
(\omega_{c2}^2 - \omega^2) - \ii W \,,
\end{equation}
where
\begin{subequations}
\label{resonances}
\begin{align}
\omega_{c1} =& \; \frac{1}{\sqrt{2}} \,
\sqrt{ k_1 + k_2 + 2 k_c + \gamma_1 \gamma_2 -
\sqrt{ L } } \,,
\\
\omega_{c2} =& \; \frac{1}{\sqrt{2}} \,
\sqrt{ k_1 + k_2 + 2 k_c + \gamma_1 \gamma_2 +
\sqrt{ L } } \,,
\\
L =& \; (k_1 - k_2)^2 + 4 k_c^2 + 2 (k_1+k_2 + 2 k_c)
\gamma_1 \gamma_2
\nonumber\\
& \; + (\gamma_1 \gamma_2)^2 \,,
\\
W =& \;
\left[ k_1 \gamma_2 + k_2 \gamma_1 + (\gamma_1 + \gamma_2) (k_c - \omega^2)
\right] \omega \,.
\end{align}
\end{subequations}

Provided that $k_c \ll k_1, k_2, | k_1 - k_2|$
(which implies non-degenerate uncoupled oscillator resonance
frequencies), one can Taylor-expand in $k_c$ as follows,
\begin{subequations}
\label{upshift}
\begin{align}
\omega_{c1} =& \; \sqrt{k_1} + \frac{k_c}{2 \sqrt{k_1}} -
\frac{(3 k_1 + k_2) k_c^2 + 4 k_1^2 \gamma_1 \gamma_2}%
{8 \, (k_1)^{3/2} \, (k_2 - k_1)} + \dots \,,
\\
\omega_{c2} =& \; \sqrt{k_2} + \frac{k_c}{2 \sqrt{k_2}} +
\frac{(3 k_2 + k_1) k_c^2 + 4 k_2^2 \gamma_1 \gamma_2}%
{8 \, (k_2)^{3/2} \, (k_2 - k_1)} + \dots \,.
\end{align}
\end{subequations}
In obtaining these expressions we have
assumed that $k_c \sim \gamma_1 \sim \gamma_2$
are small parameters, and we have
expanded up to second order in these three
parameters. Provided that $W$
is small near the resonances (it is proportional
to the width parameters $\gamma_1$ and $\gamma_2$),
we can assume that the shifted resonance
frequencies $\omega_{c1}$ and $\omega_{c2}$
approximate the resonance frequencies of the
coupled system reasonably well.

The imaginary part of the signal at the first resonance frequency
can be obtained analytically as follows,
\begin{equation}
\label{ImP1}
{\rm Im} [\chi(\omega_{c1})] =
\frac{a_2 k_1 + a_1 k_2 + (a_1 + a_2) (2 k_c - \omega_{c1}^2)}%
  {\omega_{c1}[k_2 \gamma_1 +
    k_1 \gamma_2 + (\gamma_1 + \gamma_2) (k_c - \omega_{c1}^2)]} \, .
\end{equation}
At the second resonance frequency, one finds
\begin{equation}
\label{ImP2}
{\rm Im} [\chi(\omega_{c2})] =
\frac{a_2 k_1 + a_1 k_2 + (a_1 + a_2) (2 k_c - \omega_{c2}^2)}%
  {\omega_{c2}[k_2 \gamma_1 +
    k_1 \gamma_2 + (\gamma_1 + \gamma_2) (k_c - \omega_{c2}^2)]} \,.
\end{equation}

\begin{figure}
\begin{center}
\includegraphics[width=0.91\linewidth]{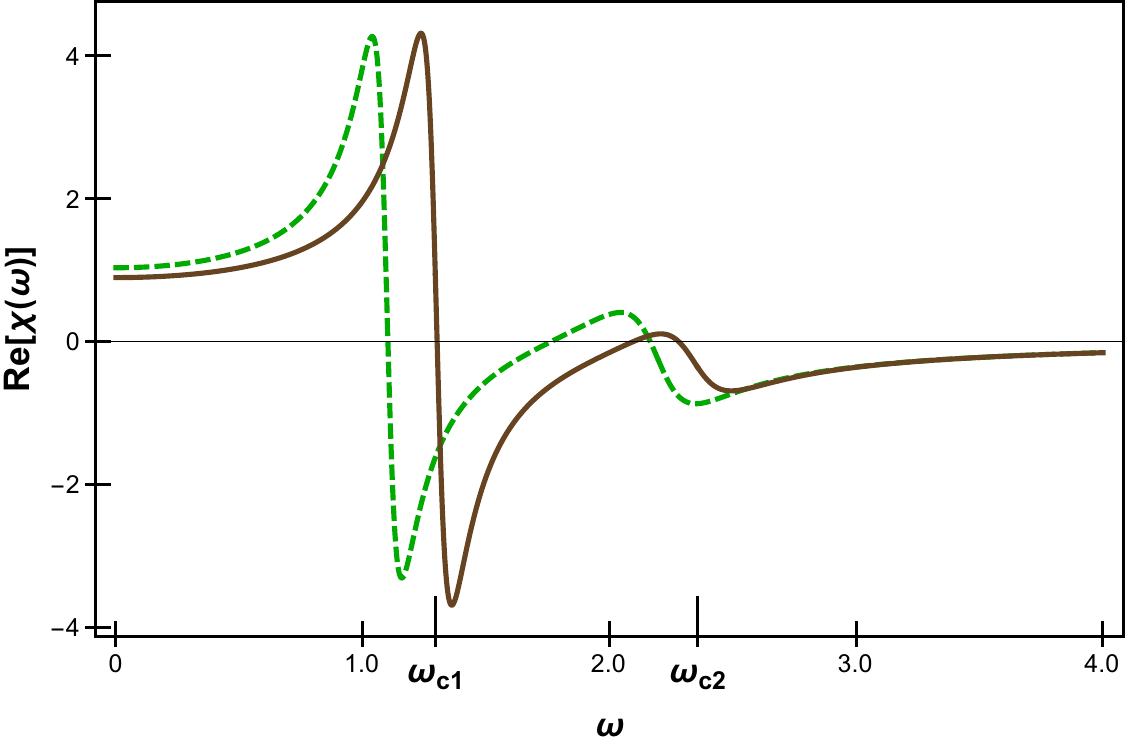}
\caption{\label{fig2}
Real part $\Re[\chi(\omega)]$ of the susceptibility
for the model problem given in Eqs.~\eqref{P2},~\eqref{params1}
and~\eqref{eps}.
The resonance frequencies of the coupled
system (solid brown curve) are shifted toward higher frequencies
in comparison to the resonance frequencies of the
uncoupled system (dashed green curve), consistent
with Eqs.~\eqref{resonances} and~\eqref{upshift}.
}
\end{center}
\end{figure}

\begin{figure}
\includegraphics[width=0.91\linewidth]{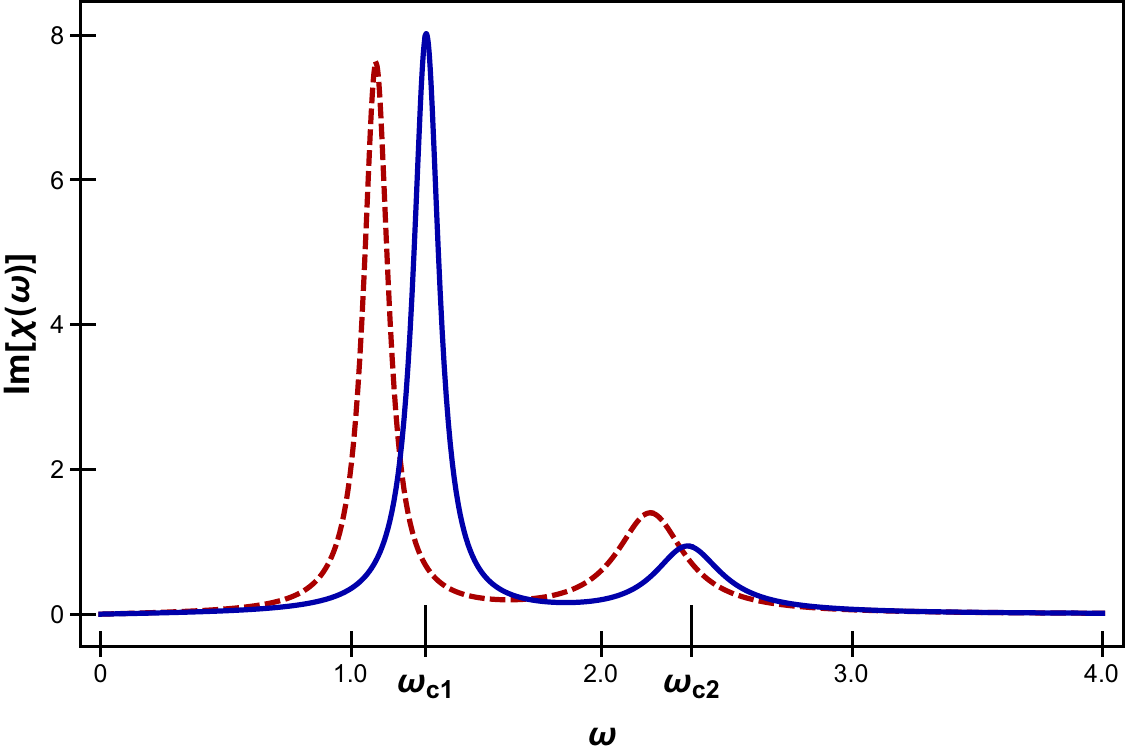}
\caption{\label{fig3}
Same as Fig.~\ref{fig2}, but for the imaginary
part of the polarization $\Im[\chi(\omega)]$,
for the model problem given in Eqs.~\eqref{params1}
and~\eqref{eps}.
The resonance frequencies of the coupled
system (solid blue curve) are shifted toward higher frequencies
in comparison to the resonance frequencies of the
uncoupled system (dashed red curve)
[see Eqs.~\eqref{resonances} and~\eqref{upshift}].
}
\end{figure}

\begin{figure}
\includegraphics[width=0.91\linewidth]{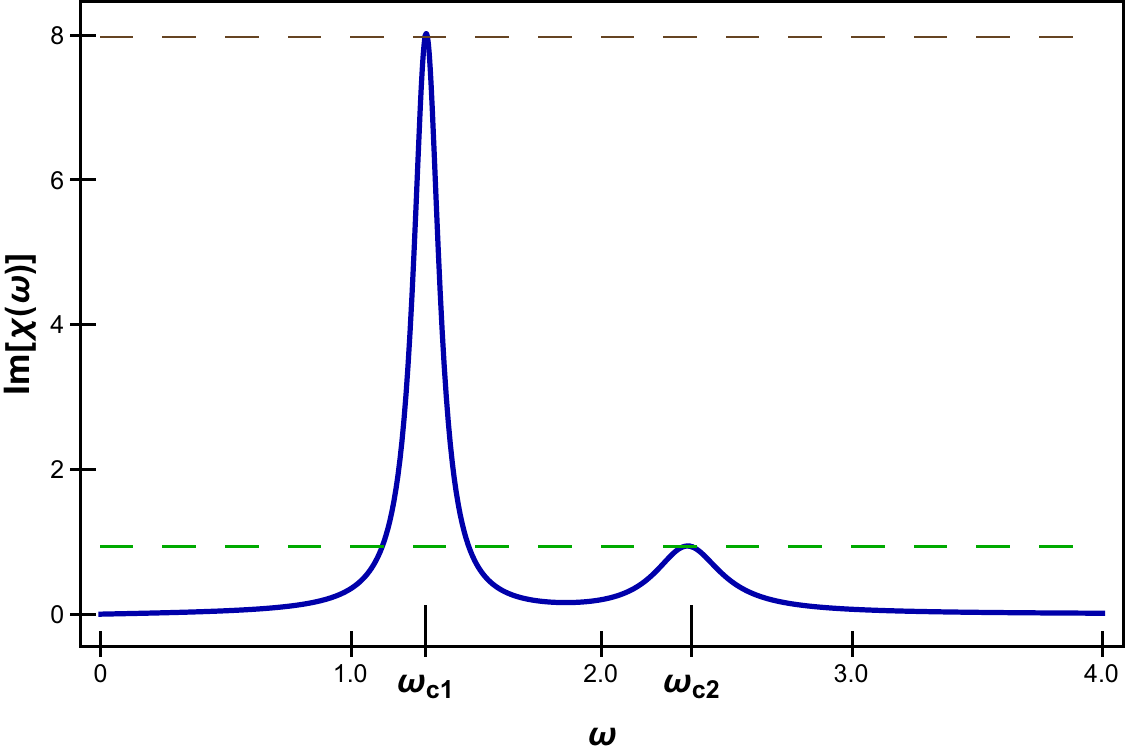}
\caption{\label{fig4}
Imaginary part $\Im[\chi(\omega)]$ for 
the model problem given in Eqs.~\eqref{params1}
and~\eqref{eps}, indicating the
resonance positions~\eqref{resonances}
and the precise values of the
maxima of the imaginary parts given in
Eqs.~\eqref{ImP1} and~\eqref{ImP2}.
The upper dashed line designates $\Im[\chi(\omega_{c1})]$,
while the lower dashed line designates $\Im[\chi(\omega_{c2})]$.}
\end{figure}

%
%
\subsection{Exact Decomposition}

To illustrate the behavior of the coupled system
we now consider a numerical example with the following parameters:
\begin{subequations}
\label{params1}
\begin{align}
\quad a_1 =& \; a_2 = 1 \,, \quad k_1 = 1.21 \,,
\quad k_2 = 4.84 \,,
\\[0.1133ex]
\gamma_1 =& \; 0.12 \,, \quad \gamma_2 = 0.33 \,.
\end{align}
We shall contrast the cases
\begin{equation}
k_c = 0.57 \qquad \mbox{versus} \qquad
k_c = 0 \,.
\end{equation}
\end{subequations}

Based on Eq.~\eqref{resonances},
one finds the following coupled resonance frequencies
for $k_c = 0.57$,
\begin{equation}
\omega_{c1} = 1.29431 \,, \quad
\omega_{c2} = 2.35677 \,,
\end{equation}
while for $k_c = 0$, the frequencies
are equal to the uncoupled ones, namely, 
$\omega_{10} = 1.1$ and $\omega_{20} = 2.2$.
In Figs.~\ref{fig2} and~\ref{fig3}, we plot
the real and imaginary parts of the
susceptibility $\chi(\omega)$
defined in Eq.~\eqref{P2},
which is related to the dielectric function,
\begin{equation}
\label{eps}
\epsilon(\omega) = 1 + \chi(\omega) \,.
\end{equation}
As evident from Figs.~\ref{fig2} and~\ref{fig3}, both resonance
frequencies of the coupled system are higher than the
corresponding one of the uncoupled system.
This is consistent with Eq.~\eqref{upshift}.
The imaginary parts of $\chi(\omega)$ at the resonances 
[see Eqs.~(\ref{ImP1}) and (\ref{ImP2})]
are explicitly indicated in Fig. \ref{fig4}.

\begin{figure}
\includegraphics[width=0.91\linewidth]{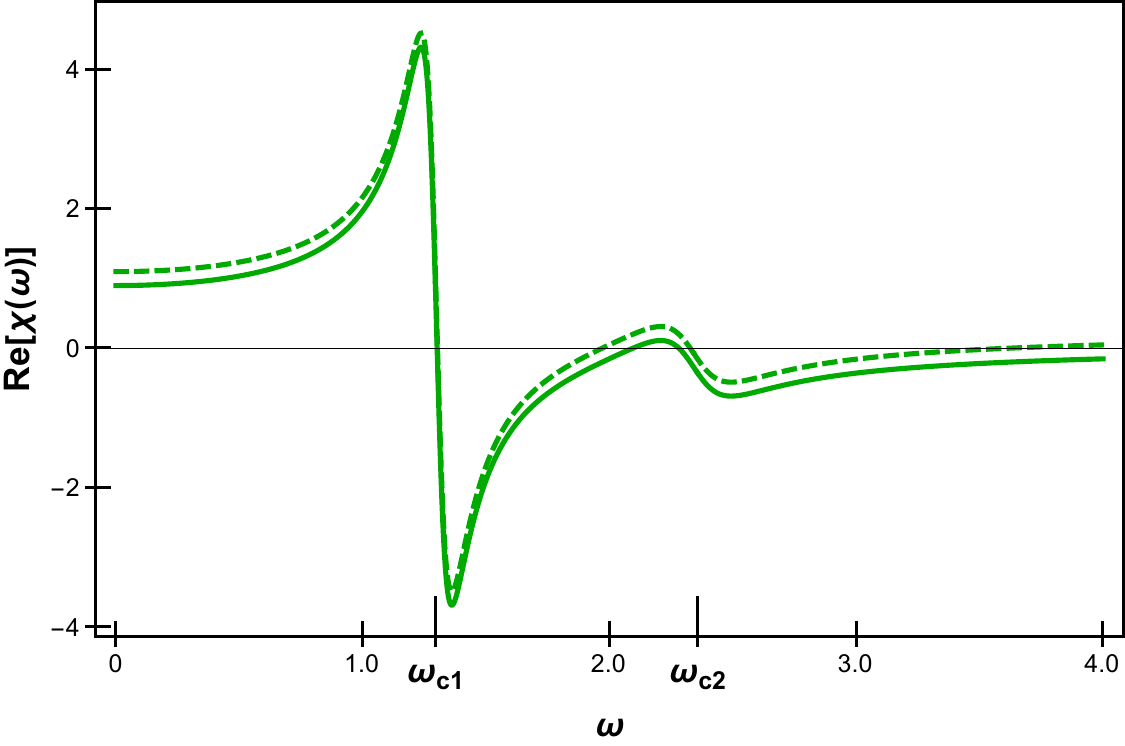}
\caption{\label{fig5}
Illustration of the exact equality
of the real part $\Re[\chi(\omega)]$ defined in Eq.~\eqref{P2} (solid curve)
and the Lorentz--Dirac (generalized
Sellmeier) representation given in Eq.~\eqref{exact} (dashed line),
for the parameters given in Eqs.~\eqref{params1}
and~\eqref{result1}.
The Lorentz--Dirac representation
has the analytic structure of Eq.~\eqref{master}.
An offset of $+0.2$ is applied to the
representation~\eqref{exact}
in order to make the two curves  visually discernible.}
\end{figure}

We now report on a surprising observation:
On the basis of a relatively
involved partial-fraction
decomposition (see Appendix~\ref{appa}),
one finds the that $\chi(\omega)$ can be {\em identically}
expressed as follows,
\begin{equation}
\label{exact}
\chi(\omega) =
\dfrac{ \tilde a_1 \, [ (\tilde\omega_1)^2 - \ii {\tilde\gamma}'_1 \omega ] }%
{ (\tilde\omega_1)^2 - \omega^2 - \ii \, \tilde\gamma_1 \, \omega}
+ \dfrac{ \tilde a_2 \, [ (\tilde\omega_2)^2 - \ii {\tilde\gamma}'_2 \omega ] }%
{ (\tilde\omega_2)^2 - \omega^2 - \ii \, \tilde\gamma_2 \, \omega} \,,
\end{equation}
with the parameters
\begin{subequations}
\label{result1}
\begin{align}
\tilde a_1 =& \; 0.767\,032\,409\,078  \,, \;
\tilde\omega_1 = 1.301\,169\,279\,269 \,,
\\[0.1133ex]
{\tilde\gamma}'_1 =& \;
-2.064\,444\,521\,508 \times 10^{-2} \,,
\\[0.1133ex]
{\tilde\gamma}_1 =& \; 0.124\,833\,194\,922 \,,
\\[0.1133ex]
\tilde a_2 =& \; 0.128\,194\,836\,781 \,,  \;
\tilde\omega_2 = 2.344\,347\,861\,458 \,,
\\[0.1133ex]
{\tilde\gamma}'_2 =& \; 0.123\,522\,592\,212 \,,  \;
{\tilde\gamma}_2 = 0.325\,166\,805\,078 \,.
\end{align}
\end{subequations}
The result~\eqref{exact} precisely 
has the structure of Eq.~\eqref{master}
and provides for an exact decomposition 
of the response of the coupled oscillators
in terms of resonators with complex oscillator
strengths.
The fact that the susceptibility $\chi(\omega)$
of the coupled oscillators can be expressed
in the form~\eqref{exact} becomes understandable
if one observes that the denominator $\calD$ defined
in Eq.~\eqref{calD} is a fourth-degree polynomial
in $\omega$ and thus has four roots
in the complex plane, with negative imaginary
parts for the roots
$\pm \sqrt{ (\tilde\omega_1)^2 - (\tilde\gamma_1)^2/4}
- \ii \tilde \gamma_1/2$ and
$\pm \sqrt{ (\tilde\omega_2)^2 - (\tilde\gamma_2)^2/4}
- \ii \tilde \gamma_2/2$.
One notes that numerically,
$\tilde\omega_1 \approx \omega_{c1}$ and
$\tilde\omega_2 \approx \omega_{c2}$,
but there is no equality (see also Appendix~\ref{appa}).
The fact that ${\tilde\gamma}'_1$ is negative
confirms that coupled oscillators may introduce
what would be referred to as a
negative radiation-reaction term in
the sense of the considerations of
Appendix A.2 of Ref.~\cite{MoEtAl2022}.
The representation~\eqref{exact}
precisely has the structure of Eq.~\eqref{master}.
In Fig.~\ref{fig5},
the equality of the expressions
in Eq.~\eqref{P2} and in~\eqref{exact}
is represented visually.

\begin{figure}
\includegraphics[width=0.91\linewidth]{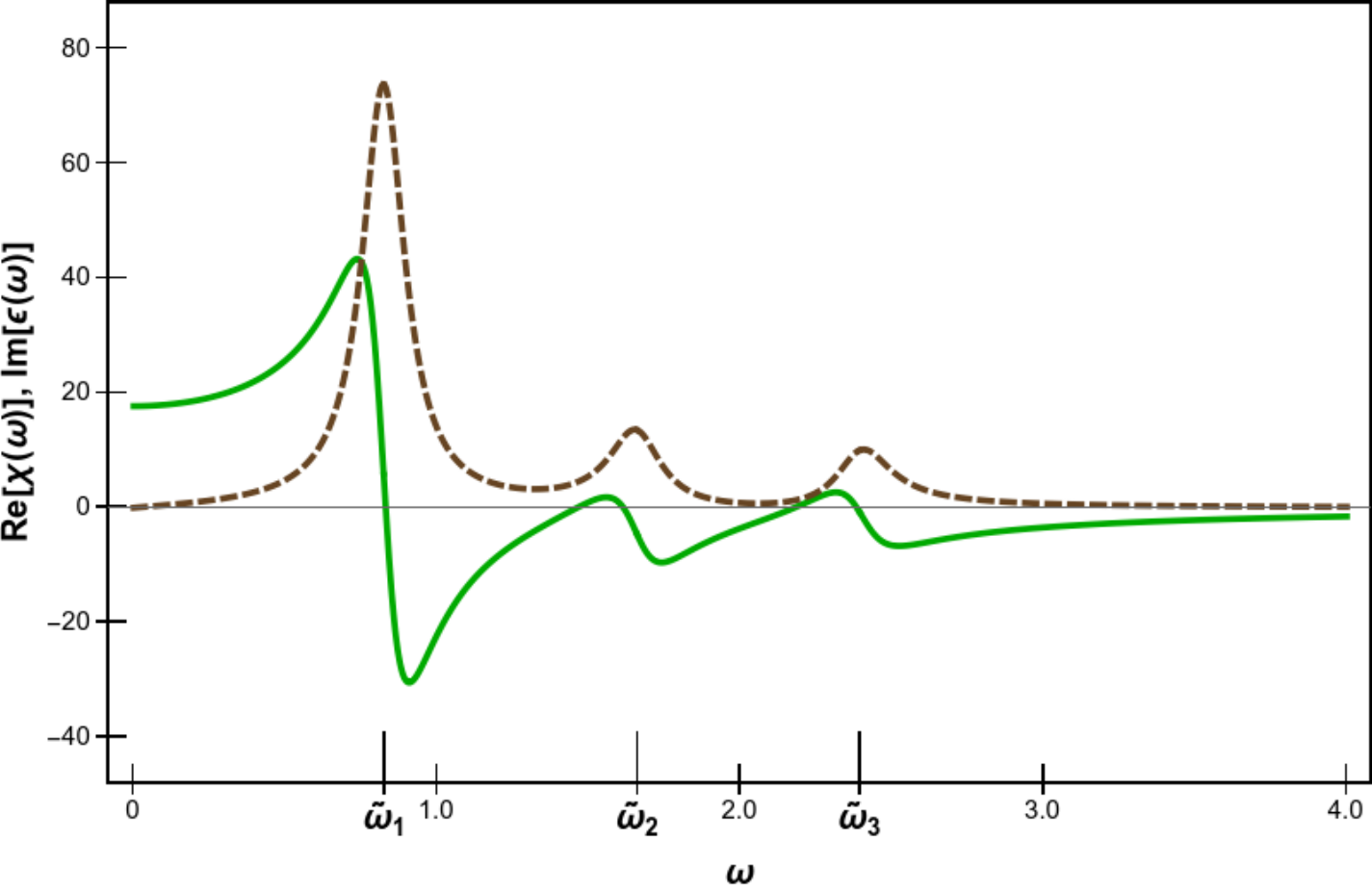}
\caption{\label{fig6}
Real and imaginary parts of $\chi(\omega)$ from Eq.~\eqref{exact3osc}
for three coupled oscillators, with parameters given in
Eqs.~\eqref{params3osc} and~\eqref{result3osc}.
The solid curve shows the real part $\Re[\chi(\omega)]$,
while the dashed curve depicts the imaginary part $\Im[\chi(\omega)]$.}
\end{figure}

%
%
\section{Three Coupled Oscillators}
\label{sec3}

We generalize the considerations of Sec.~\ref{sec2}
to three coupled oscillators, with
coordinates $x_1$, $x_2$, and $x_3$ (see Fig.~\ref{fig1}).
The equations of motion are obtained as follows,
\begin{subequations}
\begin{align}
\frac{\dd^2 x_1}{\dd t^2} =& \;
 - \gamma_1 \frac{\dd x_1}{\dd t} - k_1 x_1 + k_{c1} \, (x_2-x_1) \,,
\\
\frac{\dd^2 x_2}{\dd t^2} =& \;
 - \gamma_2 \frac{\dd x_2}{\dd t} + k_{c1} \, (x_1-x_2) + k_{c2} \, (x_3-x_2) \,,
\\
\frac{\dd^2 x_3}{\dd t^2} =& \;
 - \gamma_3 \frac{\dd x_3}{\dd t} - k_2 x_3 + k_{c2} \, (x_2-x_3) \,.
\end{align}
\end{subequations}
In frequency space, one has
$\bbmM(\omega) \cdot \vec x(\omega) = \bbmA \cdot \vec E(\omega)$,
\begin{align}
& \bbmM \cdot
\left( \begin{array}{cc}
x_1(\omega) \\ x_2(\omega) \\ x_3(\omega) \end{array} \right)
= \bbmA \cdot
\left( \begin{array}{cc}
E_1(\omega) \\ E_2(\omega) \\ E_3(\omega) \end{array} \right) \,,
\\[0.1133ex]
& \bbmM = \left( \begin{array}{ccc}
\bbmM_{11} & \bbmM_{12} & \bbmM_{13} \\
\bbmM_{21} & \bbmM_{22} & \bbmM_{23} \\
\bbmM_{31} & \bbmM_{32} & \bbmM_{33} \\
\end{array} \right) \,.
\end{align}
The $\bbmA$ matrix is obtained as follows,
\begin{equation}
\bbmA = {\rm diag}(a_1, a_2, a_3) \,.
\end{equation}
The nonvanishing elements of the $\bbmM$ matrix are easily
obtained,
\begin{subequations}
\begin{align}
\bbmM_{11} =& \; k_1 + k_{c1} - \ii \, \gamma_1 \, \omega - \omega^2 \,,
\\[0.1133ex]
\bbmM_{12} =& \; - k_{c1} = \bbmM_{21} \,,
\\[0.1133ex]
\bbmM_{22} =& \; k_{c1} + k_{c2} - \ii \gamma_2 \omega - \omega^2 \,,
\\[0.1133ex]
\bbmM_{23} =& \; - k_{c2} = \bbmM_{32} \,,
\\[0.1133ex]
\bbmM_{33} =& \; k_2 + k_{c2} - \ii \gamma_3 \omega - \omega^2 \,,
\end{align}
\end{subequations}
while $\bbmM_{13} = \bbmM_{31} = 0$.
Let us define $\bbmX = \bbmM^{-1} \cdot \bbmA$.
In analogy to Eq.~\eqref{P2},
we obtain, for $E_1(\omega) = E_2(\omega) = E_3(\omega) = E(\omega)$,
the susceptibility as
\begin{align}
\label{P3}
\chi(\omega) \equiv & \;
{\rm Tr} \, \left[ \bbmX \cdot
\left( \begin{array}{c}
1 \\ 1 \\ 1
\end{array} \right) \right] =
\sum_{i,j=1}^3 \bbmX_{ij}
=
\frac{ Q(\omega) }{ \det(\bbmM) } \,,
\end{align}
where $\det(\bbmM)$ is the determinant of the
$\bbmM$ matrix and $Q(\omega)$ constitutes a sixth-degree polynomial
in $\omega$, which is not explicitly given here.
Analytic formulas analogous to
those presented in Appendix~\ref{appa} are harder to obtain
than for the case of two coupled
oscillators, in view of the fact that the
determinant $\det( \bbmM )$, for three as opposed to 
two coupled oscillators, constitutes
a sixth-degree polynomial in $\omega$,
for which there are in general no known analytic solutions.

We thus consider the numerical example
\begin{subequations}
\label{params3osc}
\begin{align}
k_1 =& \; 0.5 \,, \quad k_{c1} = 1.3 \,, \quad
k_{c2} = 1.8 \,, \quad k_2 = 2.5 \,,
\\[0.1133ex]
a_1 =& \; 1.5 \,, \quad a_2 = 0.5 \,, \quad
a_3 = 3.0 \,,
\\[0.1133ex]
\gamma_1 =& \; 0.2 \,, \quad
\gamma_2 = 0.1 \,, \quad
\gamma_3 = 0.3 \,. \quad
\end{align}
\end{subequations}
Via a partial-fraction decomposition
analogous to the one outlined in Appendix~\ref{appa}
(but more complicated because
of the more complex nature of the
entire expression encountered for
three oscillators), one arrives at the
result
\begin{equation}
\label{exact3osc}
\chi(\omega) = \sum_{k=1}^3
\dfrac{ \tilde a_k \, [ (\tilde\omega_k)^2 - \ii {\tilde\gamma}'_k \omega ] }%
{ (\tilde\omega_k)^2 - \omega^2 - \ii \, \tilde\gamma_k \, \omega} \,,
\end{equation}
with the parameters
\begin{subequations}
\label{result3osc}
\begin{align}
\tilde a_1 =& \; 15.139\,958\,453 \,, \, \quad \tilde\omega_1 = 0.828\,692\,566 \,,
\\[0.1133ex]
{\tilde\gamma}'_1 =& \; -1.337\,923\,149 \times 10^{-2} \,,
\\[0.1133ex]
{\tilde\gamma}_1 =& \; 1.713\,373\,478 \times 10^{-2}  \,,
\\[0.1133ex]
\tilde a_2 =& \; 1.642\,138\,137 \,, \, \quad \tilde\omega_2 = 1.662\,516\,990 \,,
\\[0.1133ex]
{\tilde\gamma}'_2 =& \; -2.297\,066\,450 \times 10^{-2} \,,
\\[0.1133ex]
{\tilde\gamma}_2 =& \; 2.089\,767\,725 \times 10^{-1}  \,,
\\[0.1133ex]
\tilde a_3 =& \; 0.919\,509\,649 \,, \, \quad \tilde\omega_3 = 2.395\,819\,425 \,,
\\[0.1133ex]
{\tilde\gamma}'_3 =& \; 6.305\,219\,871 \times 10^{-1} \,,
\\[0.1133ex]
{\tilde\gamma}_3 =& \; 2.196\,858\,795 \times 10^{-1} \,.
\end{align}
\end{subequations}
Again, two of the $\gamma'$ parameters are negative
(${\tilde\gamma}'_1$ and ${\tilde\gamma}'_2$),
and thus, the possibility of explaining the
presence of negative $\gamma'$ due to the coupling
of resonances is confirmed (see Ref.~\cite{Je2024multipole}
for an application to $\alpha$-quartz).
The spectrum~\eqref{exact3osc}, for the parameters
given in Eq.~\eqref{params3osc},
is shown in Fig.~\ref{fig6}.

%
%
\section{Coupled Damped Oscillators with Radiative Reaction}
\label{sec4}

At this point, both questions raised near the beginning of
Sec.~\ref{sec2} have been answered affirmatively.
There is one more point to address.
Namely, in the course of the investigations~\cite{MoEtAl2022},
a situation was encountered where
the first resonance of the Lorentz--Dirac type,
in a model of the form~\eqref{master}
with $k_{\rm max} = 2$,
when taken alone, would
induce a small negative imaginary
part in $\epsilon(\omega)$ for
small $\omega$, while the sum
over both resonances does not suffer from
this problem.

A third question thus emerges: {\em (iii)}
Is it possible to find a model system of two coupled
oscillators, with radiative reaction,
which generates a negative imaginary part
from the first of two resonances,
while the imaginary part for the
sum of both resonances remains positive?

Let us first observe that, with the
inclusion of radiative reaction,
the system~\eqref{eqsys} of equations
is modified to read
\begin{align}
\frac{\dd^2 x_1}{\dd t^2} =& \;
- \gamma_1 \, \frac{\dd x_1}{\dd t}
- (k_1 + k_c) \, x_1 + k_c \, x_2 + \calF_1(t)  \,,
\\[0.1133ex]
\frac{\dd^2 x_2}{\dd t^2} =& \;
- \gamma_2 \, \frac{\dd x_2}{\dd t}
- (k_2 + k_c) \, x_2 + k_c \, x_1 + \calF_2(t) \,.
\end{align}
In frequency space this is equivalent to the matrix equation
$\bbmM \cdot \vec x(\omega) = \vec \calF(\omega)$,
where the elements of the
$\bbmM$ matrix are given in Eq.~\eqref{M}.
Here, $\calF$ is a generalized force, which
includes radiative reaction (for which there
is no classical analogue).
With radiation-reaction included,
$\vec \calF(\omega)$ takes the form
\begin{equation}
\left( \begin{array}{cc}
\calF_1 \\ \calF_2 \end{array} \right)
=
\left( \begin{array}{cc}
R_{11} & R_{12} \\
R_{21} & R_{22} \\
\end{array} \right) \cdot
\left( \begin{array}{cc}
E_1 \\ E_2 \end{array} \right) \,,
\end{equation}
where
\begin{subequations}
\begin{align}
R_{11} =& \; a_1 (k_1 - \ii \gamma'_1 \omega) \,,
\\[0.1133ex]
R_{12} =& \; 0 = R_{21} \,,
\\[0.1133ex]
R_{22} =& \;  a_2 ( k_2 - \ii \gamma'_2 \omega ) \,.
\end{align}
\end{subequations}
We reemphasize that there is no classical analogue
for radiative reaction. One thus has the relation
\begin{equation}
\left( \begin{array}{cc}
x_1(\omega) \\ x_2(\omega) \end{array} \right) =
\bbmX \cdot
\left( \begin{array}{cc}
E_1(\omega) \\ E_2(\omega) \end{array} \right) \,,
\end{equation}
where $E_1$ and $E_2$ are the electric fields
driving the two oscillators.
The elements of $\bbmX = \bbmM^{-1} \cdot \bbmR$ are
as follows,
\begin{subequations}
\begin{align}\label{40a}
\bbmX_{11} =& \; \frac{ \calN_1 }{ \calD } \,,
\quad
\bbmX_{12} = \frac{a_2 k_c (k_2 - \ii \gamma'_2 \omega) }{\calD } \,,
\\[0.1133ex]
\bbmX_{22} =& \; \frac{a_1 k_c (k_1 - \ii \gamma'_1 \omega) }{\calD } \,,
\quad
\bbmX_{22} = \frac{ \calN_2 }{ \calD } \,,\label{40b}
\end{align}
where
\begin{align}
\calN_1 =& \;
a_1 (k_1 - \ii \gamma'_1 \omega) \,
(k_2 + k_c - \omega^2 - \ii \gamma_2 \omega) \,,
\\[0.1133ex]
\calN_2 =& \; a_2 (k_2 - \ii \gamma'_2 \omega)
(k_1 + k_c - \omega^2 - \ii \gamma_1 \omega ) \,.
\end{align}
\end{subequations}
The denominator in Eqs. (\ref{40a}) and (\ref{40b}) has the
structure $\calD = \det(\bbmM)$, where $\bbmM$
is given in Eq.~\eqref{M}.

We consider the following example parameters:
\begin{subequations}
\begin{align}
\quad a_1 =& \; a_2 = 1 \,, \;\; k_c = 0.4 \,, \;\;
k_1 = 1.2 \,,
\\[0.1133ex]
k_2 =& \; 2.4 \,, \quad
\gamma_1 = 0.12 \,, \quad
\gamma_2 = 0.2 \,,
\\[0.1133ex]
\gamma'_1 =& \; 0.2 \,, \qquad \gamma'_2 = 0.05 \,,
\end{align}
\end{subequations}
and the susceptibility is given by
\begin{equation}
\chi(\omega) =
\bbmX_{11} + \bbmX_{12} + \bbmX_{21} + \bbmX_{22} \,,
\end{equation}
in full analogy with Eq.~\eqref{P2}.
One finds the following, {\em exact} decomposition,
\begin{equation}
\label{exact_radreac}
\chi(\omega) =
\dfrac{ \tilde a_1 \, [ (\tilde\omega_1)^2 - \ii {\tilde\gamma}'_1 \omega ] }%
{ (\tilde\omega_1)^2 - \omega^2 - \ii \, \tilde\gamma_1 \, \omega}
+ \dfrac{ \tilde a_2 \, [ (\tilde\omega_2)^2 - \ii {\tilde\gamma}'_2 \omega ] }%
{ (\tilde\omega_2)^2 - \omega^2 - \ii \, \tilde\gamma_2 \, \omega} \,,
\end{equation}
with the parameters
\begin{subequations}
\label{example_radreac}
\begin{align}
\tilde a_1 =& \; 1.55425744 \, \quad \tilde\omega_1 = 1.216\,304\,356 \,,
\\[0.1133ex]
{\tilde\gamma}'_1 =& \; 0.128\,975\,503 \,, \quad
{\tilde\gamma}_1 = 0.126\,738\,002  \,,
\\[0.1133ex]
\tilde a_2 =& \; 0.445\,742\,559 \,, \quad \tilde\omega_2 = 1.708\,832\,956 \,,
\\[0.1133ex]
{\tilde\gamma}'_2 =& \; 0.111\,137\,838  \,, \quad
{\tilde\gamma}_2 = 0.193\,261\,997   \,.
\end{align}
\end{subequations}
In view of the inequality ${\tilde\gamma}'_1  > {\tilde\gamma}_1$,
the first resonance term creates a small negative
imaginary part for small $\omega$. Numerically,
one verifies that
\begin{equation}
\label{neg}
\left.
\frac{\partial}{\partial \omega}
\dfrac{ \tilde a_1 \, [ (\tilde\omega_1)^2 - \ii {\tilde\gamma}'_1 \omega ] }%
{ (\tilde\omega_1)^2 - \omega^2 - \ii \, \tilde\gamma_1 \, \omega}
\right|_{\omega = 0} = -\ii \, 2.3507 \times 10^{-3} \,.
\end{equation}
By contrast, with $\chi$ given in Eq.~\eqref{exact_radreac},
one has the following numerical result for the
derivative at $\omega$, for the full susceptibility $\chi$,
\begin{equation}
\label{pos}
\left.
\frac{\partial}{\partial \omega}
\chi(\omega)
\right|_{\omega = 0} = +\ii \, 1.0185 \times 10^{-2} \,.
\end{equation}
Because $\Im[\chi(\omega = 0)]=0$, the derivative
at $\omega = 0$ indicates the presence or
absence of a negative imaginary part for
small driving frequency $\omega$.
One concludes that it is easily possible
to devise a model problem of two coupled
oscillators (which includes radiative damping),
where one term in the decomposition~\eqref{example_radreac}
generates a spurious negative imaginary part~\eqref{neg}
of the dielectric function $\epsilon = 1 + \chi$
for small $\omega$, while the full susceptibility
$\chi$ has a positive imaginary part [see Eq.~\eqref{pos}],
in accordance with the causality principle,
and, from a different point of view,
the second law of thermodynamics.

%
%
\section{Back-reaction and Coupling}
\label{sec5}

It is very well known that, if one considers
the response of dense materials to
incident electromagnetic radiation, the
back-reaction of the emitted radiation onto the
constituents of the solid needs to be
considered. On the classical level
this is manifest in the derivation of the
Clausius--Mossotti equation (see, {\em e.g.},
Refs.~\cite{AsMe1976,Ha1983} or
Sec.~6.4.4 of Ref.~\cite{Je2017book}). Namely,
in order to derive the Clausius--Mossotti equation,
one considers an external electric field
which orients the dipoles inside the solid.
However, the oriented dipoles, in turn,
generate an additional electric field
which needs to be added 
to the ambient external electric field which
led to the orientation of the dipoles in the first
place. Relating the polarization generated
with the local electric field, which takes the
field generated by all other polarized dipoles
in the solid into account, one obtains the
Clausius--Mossotti equation~\cite{AsMe1976,Ha1983,Je2017book}.
For a dense material, it is
insufficient to consider individual, isolated
excitations like in a dilute gas.

Now, it is well known that the oscillators in a solid
are coupled via the crystal lattice.
For low-energy resonances, this is clear
because the phonons themselves are
quantized lattice vibrations~\cite{AsMe1976}.
In Ref.~\cite{FrLi1995}, a Hamiltonian
has been explored which describes the coupling
of the electrons in the crystal lattice
to the (quantized) phonon field. The
structure of the phonon and electron-phonon
coupling terms is reminiscent of coupled
harmonic oscillators.

For higher-energy resonances (with an energy
higher than vibrational
excitations, {\em e.g.}, optical resonances), one needs to
consider the back-reaction of the substantial
polarization field (namely, the additional electric field
due to all the polarized dipoles inside the crystal)
onto a specific reference atom,
locally within the lattice (this back-reaction
effect is incorporated into
the Clausius--Mossotti equation on the
classical level).
In the optical regime, the back-reaction is
primarily of electromagnetic origin.
In Ref.~\cite{FlEtAl2019}, this back-reaction mechanism
is explored on the level of quantized fields:
The Hamiltonian considered in Ref.~\cite{FlEtAl2019}
contains a photon term with a dipole
coupling to the Kohn--Sham orbitals
obtained from density-functional theory~\cite{HoKo1964,KoSh1965},
via the total dipole moment $\vec R(t)$ defined in
the text following Eq.~(6) of Ref.~\cite{FlEtAl2019}.
In turn, the photon field enters the
mean-field exchange-correlation potential
[see Eqs.~(5) and (6) of Ref.~\cite{FlEtAl2019}].
The end result is that in Ref.~\cite{FlEtAl2019}, the authors obtain quantized
self-consistent field equations,
which take the back-reaction of matter onto the
photon field into account (and vice versa).
In the text following Eq.~(8) of Ref.~\cite{FlEtAl2019},
explicit reference is made to an
analogy of the obtained self-consistent equations to
coupled harmonic oscillators.

In both cases considered above
(lattice vibrations and optical resonance), the oscillations due to
lattice vibrations, and those due to electronic excitations
of the atoms in the crystal lattice, are coupled.
These observations suggest that,
in order to generalize the treatment sketched
here for classical oscillators to the
fully quantized formalism,
the analogies of the quantized systems to
coupled oscillators pointed out in Refs.~\cite{FrLi1995,FlEtAl2019}
should be very useful.

%
%
\section{Conclusions}
\label{sec6}

In this paper, we have raised three questions (see Sec.~\ref{sec2}
and~\ref{sec4}) about the suitability of the Lorentz--Dirac
model~\eqref{master}, which could be answered affirmatively in view of the
analogies of quantum resonances in solids with coupled oscillators discussed in
Sec. \ref{sec5}.

Let us recall the first question, slightly paraphrased: {\em (i)} Can the
response of coupled damped oscillators be described to good numerical accuracy
by a functional form of the Lorentz--Dirac model~\eqref{master}, that is, can
the signal of coupled damped oscillators alternatively be described by
uncoupled oscillators with a complex-valued oscillator strength?  The answer is
affirmative: in the case of two coupled oscillators [Eq.~\eqref{exact}], three
coupled oscillators [Eq.~\eqref{exact3osc}], and two coupled oscillators with
radiative reaction [Eq.~\eqref{exact_radreac}], it was possible to bring the
model polarization response of the model system (coupled oscillators) into a
form which is exactly of the Lorentz--Dirac form~\eqref{master}.

The second question was: {\em (ii)} Are there cases where the
radiative-reaction parameter $\gamma'$ becomes negative upon fitting the signal
from the coupled resonances?  The answer is again affirmative: The $\gamma'_1$
parameter in the example cases given in Eq.~\eqref{exact} (for two coupled
oscillators) and for $\gamma'_1$ and $\gamma'_2$ in Eq.~\eqref{exact3osc} (for
three coupled oscillators) are negative. These results, obtained for our model
problems, are analogous to the presence of negative $\gamma'$ parameters for
$\alpha$-quartz~\cite{Je2024multipole}.

The third question was asked in Sec.~\ref{sec4}: {\em (iii)} Is it possible to
find a model system of two coupled oscillators, with radiative reaction, which
generates a negative imaginary part from the first of two resonances, while the
imaginary part for the sum of both resonances remains positive?  Again, the
answer is yes: For the parameters given in Eq.~\eqref{example_radreac}, the
first resonance, upon the inclusion of radiative reaction, gives rise to a
negative imaginary part for the model dielectric function [see
Eq.~\eqref{neg}], while the full response of the system does not follow this
behavior [see Eq.~\eqref{pos}].

Our considerations support the suitability of the functional
form~\eqref{master} for the description of the dielectric function of solids.
In this paper, we only considered the explicit cases of two and three
oscillators.  The functional form~\mbox{\eqref{master}} is given for an
arbitrary number of oscillators.  The analytic and numerical results presented
here strongly suggest that our findings hold for more than three oscillators;
a general analytic proof, however, is beyond the scope of this paper.

We also discussed the significance of our classical coupled oscillator model,
in the sense of a possible correspondence to a
quantum system (see Sec.~\mbox{\ref{sec5}}). Our findings suggest
that generalized oscillator strengths
(not necessarily positive and real) can
occur when different subsystems are coupled and can exchange (or dissipate)
energy. Analogies to time-dependent density-functional
theory (TDDFT), when coupled self-consistently
to photons, have been explored in Ref.~\mbox{\cite{FlEtAl2019}},
with corresponding analogies to coupled oscillators.
Our findings indicate that the self-consistent inclusion of
back-reaction effects could be an essential step
in first-principles approaches to optical absorption
spectra in the presence of electron-phonon
interactions~\cite{PaGi2014,Gi2017,MoDrRa2018}.

Finally, let us point out an interesting observation: \mbox{\em a priori}, one
might assume that the Clausius--Mossotti model studied in
Ref.~\cite{MoEtAl2022} might provide for a better representation of the
response of intrinsic silicon (certainly, a dense material) than the
Lorentz--Dirac model for the dielectric function.  However, if we conclude that
the back-reaction of the coupled oscillators is already encoded in the complex
oscillator strengths, and does not necessitate to additionally invoke the
Clausius--Mossotti inspired functional form for the dielectric function studied
in Ref.~\cite{MoEtAl2022}, then it becomes immediately understandable why the
Lorentz--Dirac and Clausius--Mossotti functional forms were seen to yield
equally satisfactory representations of the updated experimental data for the
real and imaginary parts of the dielectric function of intrinsic silicon
studied in Ref.~\cite{MoEtAl2022}.

As a final remark, we observe that the model based on the functional form
(\ref{master}) turns out to be physically well motivated and grounded in a
coupled oscillator model.
The applicability of the model~\eqref{master} for solids of
technological significance ({\em e.g.}, calcium fluoride) is currently being
studied further~\cite{DaUlJe2025fluorspar,Li1980halides,Be1985,%
DaMa2002,LeEtAl2015temp,KGEtAl2017,ZhWaTh2023}.

%
%
\section*{Acknowledgments}

Stimulating conversations with 
Professor Vladimir M.~Mostepanenko are gratefully acknowledged.
T.~D.~and U.~D.~J.~were supported by NSF grant PHY--2110294.
C.~A.~U.~acknowledges support from NSF grant DMR--2149082.

\appendix

\begin{widetext}

%
%
\section{An Important Identity}
\label{appa}

If we attempt to show that the signal of coupled
oscillators, as given in Eq.~\eqref{P2},
can be represented in the Lorentz--Dirac functional
form, then we should investigate the
functional form of the common
denominator $\calD$ in closer detail
[see Eq.~\eqref{calD}].
First, we observe that since $\calD$ corresponds
to a retarded Green function (of the coupled-oscillator
system), the roots of the equation $\calD = 0$
have a negative imaginary part.
We denote the values of the roots
as $\omega = \overline\omega_i$ (with $i=1,2,3,4$),
$\overline\omega_1 = \calE_1 - \ii \tilde\gamma_1/2$,
$\overline\omega_2 = -\calE_1 - \ii \tilde\gamma_1/2$,
$\overline\omega_3 = \calE_2 - \ii \tilde\gamma_2/2$,
$\overline\omega_4 = -\calE_2 - \ii \tilde\gamma_2/2$;
all of them have a negative imaginary part.
Our initial expression given in
Eq.~\eqref{P2} therefore has the structure
\begin{equation}
G = \frac{A + \ii \, B \, \omega + C \, \omega^2}{
\left(\calE_1 - \frac{\ii}{2} \tilde\gamma_1 - \omega \right)
\left(\calE_1 + \frac{\ii}{2} \tilde\gamma_1 + \omega \right)
\left(\calE_2 - \frac{\ii}{2} \tilde\gamma_2 - \omega \right)
\left(\calE_2 + \frac{\ii}{2} \tilde\gamma_2 + \omega \right)} \,,
\end{equation}
where $A$, $B$, $C$, $\calE_1$, $\calE_2$,
$\tilde\gamma_1$ and $\tilde\gamma_2$ are constants.
By a partial-fraction decomposition, one can show that
\begin{equation}
\label{id}
G = H \, \left( \frac{A_1 + \ii B_1 \omega}{D_1} +
\frac{A_2 + \ii B_2 \omega}{D_2}  \right) \,.
\end{equation}
The parameters can be expressed as
\begin{subequations}
\begin{align}
H =& \; \frac{1}{
\left[ 4 (\calE_1 - \calE_2)^2 + (\tilde\gamma_1 - \tilde\gamma_2)^2 \right] \,
\left[ 4 (\calE_1 - \calE_2)^2 + (\tilde\gamma_1 - \tilde\gamma_2)^2 \right] } \,,
\\[0.1133ex]
A_1 =& \;
 4 \, A \, [ 4 (\calE_2^2 - \calE_1^2) + 3 \tilde\gamma_1^2 -
   4 \tilde\gamma_1 \tilde\gamma_2 + \tilde\gamma_2^2 ]
 + 4 B (4 \calE_1^2 + \tilde\gamma_1^2) (\tilde\gamma_2 - \tilde\gamma_1)
\nonumber\\[0.1133ex]
& \; + C (4 \calE_1^2 + \tilde\gamma_1^2)
  (4 (-\calE_1^2 + \calE_2^2) - \tilde\gamma_1^2 + \tilde\gamma_2^2) \,,
\\[0.1133ex]
A_2 =& \;
 4 \, A \, [ 4 (\calE_1^2 - \calE_2^2) + \tilde\gamma_1^2
   - 4 \tilde\gamma_1 \tilde\gamma_2 + 3 \tilde\gamma_2^2 ]
\nonumber\\[0.1133ex]
& \; + 4 B (4 \calE_2^2 + \tilde\gamma_2^2) (\tilde\gamma_1 - \tilde\gamma_2)
 + C (4 \calE_2^2 + \tilde\gamma_2^2) (4 (\calE_1^2 - \calE_2^2) + \tilde\gamma_1^2 - \tilde\gamma_2^2) \,,
\\[0.1133ex]
B_1 =& \;
 16 \, A \, (\tilde\gamma_2 - \tilde\gamma_1)
 + 4 \, B \, (4 (\calE_1^2 - \calE_2^2) + \tilde\gamma_1^2 - \tilde\gamma_2^2)
\nonumber\\[0.1133ex]
& \; + 4 \, C \, (4 \calE_1^2 \tilde\gamma_2 - 4 \calE_2^2 \tilde\gamma_1
   + \tilde\gamma_1 \tilde\gamma_2 (\tilde\gamma_1 - \tilde\gamma_2) ) \,,
\\[0.1133ex]
B_2 =& \;
  16 \, A \, (\tilde\gamma_1 - \tilde\gamma_2)
  + 4 \, B \, (4 (-\calE_1^2 + \calE_2^2) - \tilde\gamma_1^2 + \tilde\gamma_2^2)
\nonumber\\[0.1133ex]
& \; + 4 \, C \, (4 \calE_2^2 \tilde\gamma_1 - 4 \calE_1^2 \tilde\gamma_2
    + \tilde\gamma_1 \tilde\gamma_2 (\tilde\gamma_2 - \tilde\gamma_1)) \,,
\\[0.1133ex]
D_1 =& \;
  \calE_1^2 + \tfrac14 \, \tilde\gamma_1^2 -
   \ii \, \tilde\gamma_1 \, \omega - \omega^2 =
  \tilde\omega_1^2 - \ii \, \tilde\gamma_1 \, \omega - \omega^2 \,,
\qquad
\tilde\omega_1 = \sqrt{ \calE_1^2 + \tfrac14 \, \tilde\gamma_1^2 } \,,
\\[0.1133ex]
D_2 =& \;
  \calE_2^2 + \tfrac14 \, \tilde\gamma_2^2 -
   \ii \, \tilde\gamma_2 \, \omega - \omega^2 =
  \tilde\omega_2^2 - \ii \, \tilde\gamma_2 \, \omega - \omega^2 \,.
\qquad
  \tilde\omega_2 = \sqrt{ \calE_2^2 + \tfrac14 \, \tilde\gamma_2^2 } \,.
\end{align}
\end{subequations}
An application of the identity~\eqref{id}
leads to the result~\eqref{exact}.
One notes that the quantities $\omega_{c1}$
and $\omega_{c2}$ constitute approximations to
$\tilde\omega_1$ and $\tilde\omega_2$, but there is no
equality: Namely, at $\omega_{c1}$
and $\omega_{c2}$, the real part of the denominator $\calD$
defined in Eq.~\eqref{calD} vanishes.
By contrast, the complex roots of the entire (complex rather than real)
denominator polynomial (in $\omega$)
are denoted as $\overline\omega_i$ with $i=1,2,3,4$.

\end{widetext}


\begin{thebibliography}{10}

\bibitem{Pa1985}
E.~D. Palik, {\em \relax{Handbook of Optical Constants of Solids}} (Academic
  Press, San Diego, 1985).

\bibitem{PSEtAl2009}
T. Passerat~de Silans, I. Maurin, P. Chaves~de Souza~Segundo, S. Saltiel, M.-P.
  Gorza, M. Ducloy, D. Bloch, D. de~Sousa~Meneses, and P. Echegut, {\em
  \relax{Temperature dependence of the dielectric permittivity of CaF${}_2$,
  BaF${}_2$ and Al${}_2$O$_3$: Application to the prediction of a
  temperature-dependent van der Waals surface interaction exerted onto a
  neighbouring Cs(8P${}_{3/2}$) atom}},  J. Phys.: Condens. Matter {\bf 21},
  255902  (2009).

\bibitem{HaKo2009}
H. Haug and S.~W. Koch, {\em \relax{Quantum Theory of the Optical and
  Electronic Properties of Semiconductors}} (World Scientific, Singapore,
  2009).

\bibitem{Je2017book}
U.~D. Jentschura, {\em \relax{Advanced Classical Electrodynamics: Green
  Functions, Regularizations, Multipole Decompositions}} (World Scientific,
  Singapore, 2017).

\bibitem{MoEtAl2022}
C. Moore, C.~M. Adhikari, T. Das, L. Resch, C.~A. Ullrich, and U.~D.
  Jentschura, {\em \relax{Temperature-dependent dielectric function of
  intrinsic silicon: Analytic models and atom-surface potentials}},  Phys. Rev.
  B {\bf 106},  045202  (2022).

\bibitem{MoEtAl2024erratum}
C. Moore, C.~M. Adhikari, T. Das, L. Resch, C.~A. Ullrich, and U.~D.
  Jentschura, {\em Erratum: Temperature-dependent dielectric function of
  intrinsic silicon: Analytic models and atom-surface potentials}, submitted.

\bibitem{Ri1985}
M.~W. Ribarsky, {\em Titanium Dioxide (Ti O$_2$) (Rutile)}, in {\em Vol.~II of
  the Handbook of Optical Constants of Solids}, pp.~795--804, (E.~D.~Palik,
  Ed.), Academic Press, Boston, 1985.

\bibitem{Tr1985}
W.~J. Tropf, {\em Cubic Thallium (I) Halides}, in {\em Vol.~III of the Handbook
  of Optical Constants of Solids}, pp.~923--967, (E.~D.~Palik, Ed.), Academic
  Press, Boston, 1985.

\bibitem{PaKh1985}
E.~D. Palik and R. Khanna, {\em Sodium Nitrate (NaNO$_3$)}, in {\em Vol.~III of
  the Handbook of Optical Constants of Solids}, pp.~873--881, (E.~D.~Palik,
  Ed.), Academic Press, Boston, 1985.

\bibitem{Je2024multipole}
U.~D. Jentschura, {\em \relax{Revisiting the Divergent Multipole Expansion of
  Atom--Surface Interactions: Hydrogen and Positronium, $\alpha$--Quartz, and
  Physisorption}},  Phys. Rev. A {\bf 109},  012802  (2024).

\bibitem{JeAd2022book}
U.~D. Jentschura and G.~S. Adkins, {\em \relax{Quantum Electrodynamics: Atoms,
  Lasers and Gravity}} (World Scientific, Singapore, 2022).

\bibitem{massfootnote}
In principle, the derivations presented here could be generalized to
  oscillators with different masses $m_1$, $m_2$, and so on. This would
  somewhat complicate the derivations but would not alter our conclusions.

\bibitem{AsMe1976}
N.~W. Ashcroft and N.~D. Mermin, {\em \relax{Solid state physics}} (Saunders
  College, Fort Worth, 1976).

\bibitem{Ha1983}
J.~H. Hannay, {\em \relax{The Clausius-Mossotti equation: an alternative
  derivation}},  Eur. J. Phys. {\bf 4},  141  (1983).

\bibitem{FrLi1995}
J.~K. Freericks and E.~H. Lieb, {\em \relax{Ground state of a general
  electron-phonon Hamiltonian as a spin singlet}},  Phys. Rev. B {\bf 51},
  2812--2821  (1995).

\bibitem{FlEtAl2019}
J. Flick, D.~M. Welakuh, M. Ruggenthaler, H. Appel, and A. Rubio, {\em
  \relax{Light-Matter Response in Non--Relativistic Quantum Electrodynamics:
  Quantum Modifications of Maxwell's Equations}},  ACS Photonics {\bf 6},
  2757--2778  (2019).

\bibitem{HoKo1964}
P. Hohenberg and W. Kohn, {\em \relax{Inhomogeneous Electron Gas}},  Phys. Rev.
  {\bf 136},  B864--B871  (1964).

\bibitem{KoSh1965}
W. Kohn and L.~J. Sham, {\em \relax{Self-Consistent Equations Including
  Exchange and Correlation Effects}},  Phys. Rev. {\bf 140},  A1133--A1138
  (1965).

\bibitem{PaGi2014}
C.~E. Patrick and F. Giustino, {\em \relax{Unified theory of electron–phonon
  renormalization and phonon-assisted optical absorption}},  J. Phys.: Condens.
  Matter {\bf 26},  365503  (2014).

\bibitem{Gi2017}
F. Giustino, {\em \relax{Electron-phonon interactions from first principles}},
  Rev. Mod. Phys. {\bf 89},  015003  (2017).

\bibitem{MoDrRa2018}
B. Monserrat, C.~E. Dreyer, and K.~M. Rabe, {\em \relax{Phonon-assisted optical
  absorption in BaSnO$_3$ from first principles}},  Phys. Rev. B {\bf 97},
  104310  (2018).

\bibitem{DaUlJe2025fluorspar}
T. Das, C.~A. Ullrich, and U.~D. Jentschura, {\em Temperature--Dependent
  Dielectric Function of Calcium Difluorite (CaF$_{\bm{2}}$): Infrared and
  Ultraviolet Contributions}, in preparation (2025).

\bibitem{Li1980halides}
H.~H. Li, {\em \relax{Refractive Index of Alkaline Earth Halides and Its
  Wavelength and Temperature Derivatives}},  J. Phys. Chem. Ref. Data {\bf 9},
  161--290  (1980).

\bibitem{Be1985}
D.~F. Bezuidenhout, {\em Calcium Fluoride (CaF$_2$)}, in {\em Handbook of
  Optical Constants of Solids, Vol.~2}, pp.~815--828, (E.~D.~Palik, Ed.),
  Academic Press, Boston, 1985.

\bibitem{DaMa2002}
M. Daimon and A. Masumura, {\em \relax{High-Accuracy Measurements of the
  Refractive Index and Its Temperature Coefficient of Calcium Fluoride in a
  Wide Wavelength Range from 138 to 2326 Nm}},  Appl. Opt. {\bf 41},  5275
  (2002).

\bibitem{LeEtAl2015temp}
D.~B. Leviton, K.~H. Miller, M.~A. Quijada, and F.~U. Grupp,  in {\em
  \relax{SPIE Optical Engineering + Applications, Proceedings Volume 9578}},
  edited by R.~B. Johnson, V.~N. Mahajan, and S. Thibault (SPIE Press, San
  Diego, California, United States, 2015), pp.\ 1--12.

\bibitem{KGEtAl2017}
M.~R.~K. Kelly-Gorham, B.~M. DeVetter, C.~S. Brauer, B.~D. Cannon, S.~D.
  Burton, M. Bliss, T.~J. Johnson, and T.~L. Myers, {\em \relax{Complex
  Refractive Index Measurements for BaF{\textsubscript{2}} and
  CaF{\textsubscript{2}} via Single-Angle Infrared Reflectance Spectroscopy}},
  Opt. Mat. {\bf 72},  743--748  (2017).

\bibitem{ZhWaTh2023}
Q. Zheng, X. Wang, and D. Thompson, {\em \relax{Temperature-Dependent Optical
  Properties of Monocrystalline CaF{\textsubscript{2}},
  {{BaF}}{\textsubscript{2}}, and {{MgF}}{\textsubscript{2}}}},  Opt. Mater.
  Express {\bf 13},  2380  (2023).

\end{thebibliography}
\end{document}